\def\year{2022}\relax
\documentclass[letterpaper]{article} 
\usepackage{aaai22}  
\usepackage{times}  
\usepackage{helvet}  
\usepackage{courier}  
\usepackage[hyphens]{url}  
\usepackage{graphicx} 
\usepackage{natbib}  
\usepackage{caption} 
\DeclareCaptionStyle{ruled}{labelfont=normalfont,labelsep=colon,strut=off} 
\usepackage{algorithm}
\usepackage{algorithmic}
\usepackage{amsfonts}
\usepackage{amsmath}
\usepackage{amssymb}
\usepackage{newfloat}
\usepackage{listings}
\floatstyle{ruled}
\newfloat{listing}{tb}{lst}{}
\floatname{listing}{Listing}
\title{Multi-Agent Intention Sharing via Leader-Follower Forest}
\author{
   Zeyang Liu\textsuperscript{\rm 1}\equalcontrib,
   Lipeng Wan\textsuperscript{\rm 1}\equalcontrib,
   Xue Sui\textsuperscript{\rm 1},
   Kewu Sun\textsuperscript{\rm 2},
   Xuguang Lan\textsuperscript{\rm 1}\thanks{Corresponding author.}
}
\affiliations{
   \textsuperscript{\rm 1} Institute of Artificial Intelligence and Robotics, Xi’an Jiaotong University\\
   \textsuperscript{\rm 2} X Lab, the second academy of CASIC, Beijing 100854, China\\
   \{zeyang.liu, wanlipeng, suixue98\}@stu.xjtu.edu.cn,\\
   sun\_kewu@126.com, xglan@mail.xjtu.edu.cn
}

\begin{document}

\maketitle

\begin{abstract}
Intention sharing is crucial for efficient cooperation under partially observable environments in multi-agent reinforcement learning (MARL). However, message 
deceiving, i.e., a mismatch between the propagated intentions and the final decisions, may happen when agents change strategies simultaneously according to received 
intentions. Message deceiving leads to potential miscoordination and difficulty for policy learning. This paper proposes the leader-follower forest (LFF) to learn 
the hierarchical relationship between agents based on interdependencies, achieving one-sided intention sharing in multi-agent communication. By limiting the flowings
of intentions through directed edges, intention sharing via LFF (IS-LFF) can eliminate message deceiving effectively and achieve better coordination. In addition, a 
two-stage learning algorithm is proposed to train the forest and the agent network. We evaluate IS-LFF on multiple partially observable MARL benchmarks, and the 
experimental results show that our method outperforms state-of-the-art communication algorithms.

\end{abstract}

\section{Introduction}

Communication between agents is key to improving the efficiency of coordination by exchanging messages, particularly in complex real-world applications under 
partially observable environments, such as multi-robot exploration \cite{matignon2012coordinated} and autonomous vehicles \cite{cao2012overview}. 
Recently, learning communication protocols instead of pre-designed protocols has become an active area in multi-agent reinforcement learning \cite{hernandez2019survey}.

Previous studies mainly focus on learning when and whom to share local observations, improving coordination compared to independent learning \cite{sukhbaatar2016learning, jiang2018learning, das2019tarmac}. 
However, the learning suffers from the environment non-stationarity, which is caused by the changing of agents’ policies during learning
procedure \cite{li2021dealing}. 
Modelling others has been put forward to alleviate the non-stationarity by predicting others' behaviours or goals \cite{raileanu2018modeling,rabinowitz2018machine}.
From the perspective of communication, agents can benefit from intention sharing and know others' intentions directly instead of inference \cite{qu2020intention,kim2020communication}.
However, the agents may change their actions simultaneously after intention sharing, resulting in message deceiving, which is defined as the mismatch between
the propagated intentions and the final decisions.
Message deceiving could degrade coordination and impair learning stability. 
For instance, two agents arrive at a crossroad where both of them send their intentions ``wait'' to the other. After knowing that the other would make way, 
they simultaneously change their actions from ``wait'' to ``move''.
In this case, agents are misled into making wrong decisions based on deceptive messages, resulting in a collision.
A simple yet effective method to address message deceiving is one-sided intention sharing, i.e., only one agent has access to the other plans to do \cite{xie2020optimally},
which ensures that the intention from the agent is consistent with its final decision. 

The key to one-sided intention sharing in multi-agent systems is learning appropriate hierarchical relationships amongst agents. Recent works use attention mechanisms to 
build the relationships between agents by graph models \cite{liu2020multi, li2021deep}. However, they focus on the undirected graphs and cannot learn the leader-follower 
relationship between agents. 

\begin{figure*}[t]
    \centering
    \includegraphics[width=1.0\textwidth]{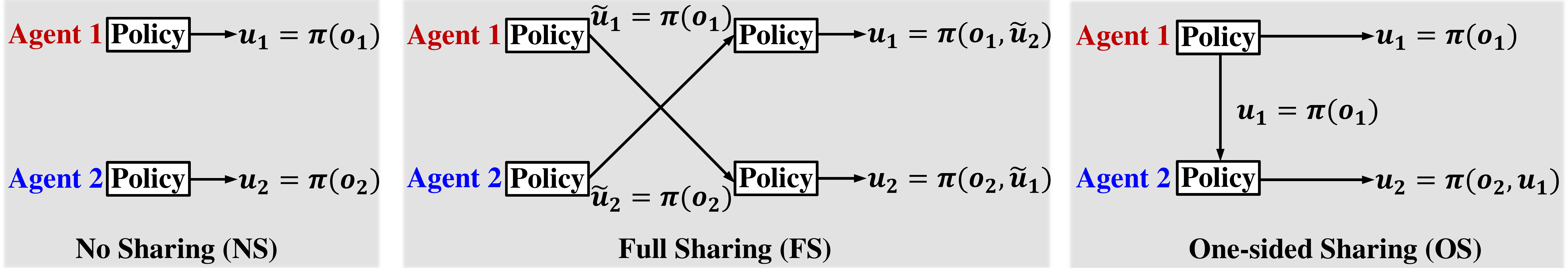} 
    \caption{Examples of intention sharing for two agents: (a) No sharing (as in VDN); (b) Full sharing. Agents share their decision information and change their policies 
    after communication; (c) One-sided sharing. Only one agent can get access to the intention information of another one.}
    \label{fig:motivation}
\end{figure*}

In this paper, we propose a new communication scheme called intention sharing via leader-follower forest (IS-LFF) that explicitly models the observation-dependent leader-follower 
relationships to constrain the flowings of intentions through directed edges. We adopt an attention module to learn the interdependencies between agents and
generate a leader-follower forest (LFF) based on Minimum Spanning Tree \cite{prim1957shortest}. LFF eliminates message deceiving and alleviates non-stationarity
in multi-agent systems. In addition, we propose a two-stage learning algorithm to train the proposed intention sharing scheme. We list the main contributions 
as follows:

\begin{itemize}
    \item By introducing directed acyclic graphs into multi-agent systems, we propose a novel intention sharing scheme based on LFF,
    ensuring one-sided intention sharing and addressing message deceiving.
    \item We adopt deep deterministic policy gradient with a communication reward for the dynamically generated LFF, and linear value decomposition for the agent
    network. Based on LFF, the joint state-action value function is decomposed to dependent value functions, alleviating environment non-stationarity in training.
    \item We evaluate our method on partially observable MARL benchmarks predator-prey and unit micromanagement tasks in StarCraft II.
    The results demonstrate that IS-LFF outperforms state-of-the-art communication algorithms and achieves better learning stability.
\end{itemize}

\section{Background}

\subsection{Decentralised Partially Observable Markov Decision Process (Dec-POMDP)}
A communication-based multi-agent fully cooperative task in the partially observable setting can be formulated as a Dec-POMDP \cite{oliehoek2016concise}, consisting of a tuple
$ G=\langle I,S,\ \Omega,O,M,F,U,P,R,\ n,\gamma\rangle $, where $ I\equiv\left\{1,\ldots,n\right\} $ describes the set of agents, $ S $ is denotes true state of
the environment, $ \mathrm{\Omega} $ denotes the set of joint observations. At each time step, due to partially observability, an agent obtains
its observation $ o\in\mathrm{\Omega} $ based on the observation function $ O\left(s,\mathbf{u}\right):S\times \mathbf{U}\rightarrow\mathrm{\Omega} $, 
where $ \mathbf{U}\equiv U^n $ is the set of joint actions. Each agent has an action-observation history $ \tau \in T \equiv (\Omega\times U)^\ast $.
The message $m\in M$ is produced by the message function $ F(\tau): T \rightarrow M $ and then propagated to others in the communication.
Each agent i chooses an action $u_i\in U$ by a stochastic policy $\pi_i\left(u_i\middle|\tau_i,m_i\right):\left(T\times  M\right)^\ast\times U\rightarrow\left[0,1\right]$,
forming a joint action $ \mathbf{u}\in\mathbf{U} $, which leads to a shared reward according to the reward function $ r(s,\mathbf{u}):S \times \mathbf{U} \rightarrow \mathbb{R}$
and a transition on the environment through the state transition function $ P\left(s^\prime\middle| s,a\right):S\times\mathbf{U}\rightarrow S$. 
The goal of the task is to learn the decentralised action-value by maximising the total expected return $ R^t=\sum_{t=0}^{T}{\gamma^tr^t} $,
where $T$ is the length of the episode and $ \gamma\in\left[0,1\right) $ is a discount factor.

\subsection{Directed acyclic graph (DAG)}
A graph is represented as $G=\langle N,E \rangle$, where $N$ is the set of nodes, and $E\subseteq\{\left(x,y\right)|\left(x,y\right)\in N^2, x\neq y\}$ is the set of edges.
A directed graph is a graph with all edges directed from one node to another, indicating a parent-child relationship between nodes.
A path is a sequence of edges, where the ending node of each edge is the starting node of the next edge in the sequence.
A DAG is a graph that has no cycles in its paths.
A root is a node with no parents, and a leaf is a node with no children.
A directed acyclic graph denotes a directed rooted tree if there is only one root node.
In a directed rooted tree, the level of a node is the length of the path to the root node. 
For node $n_i$, ancestors are defined as nodes on the path from the root to $n_i$. 
The adjacent matrix of $A$ is a $|N|\times|N|$ binary asymmetric matrix with $ A_{ij}=1 $ if $ e_{ij}\in E $, and $ A_{ij}=0 $ if $e_{ij}\notin E$.

\section{One-sided intention sharing}\label{os}

Compared with a single-agent environment, it is difficult to learn an optimal distribution of joint actions for agents
due to the partial observability and the non-stationarity in multi-agent systems.
Intention sharing is a natural way to mitigate this problem. We investigate three different intention sharing patterns, i.e., no sharing (NS), full sharing (FS), and one-sided sharing (OS).
Examples are shown in Figure \ref{fig:motivation}.
\begin{figure}[t]
    \centering 
    \includegraphics[width=1.0\columnwidth]{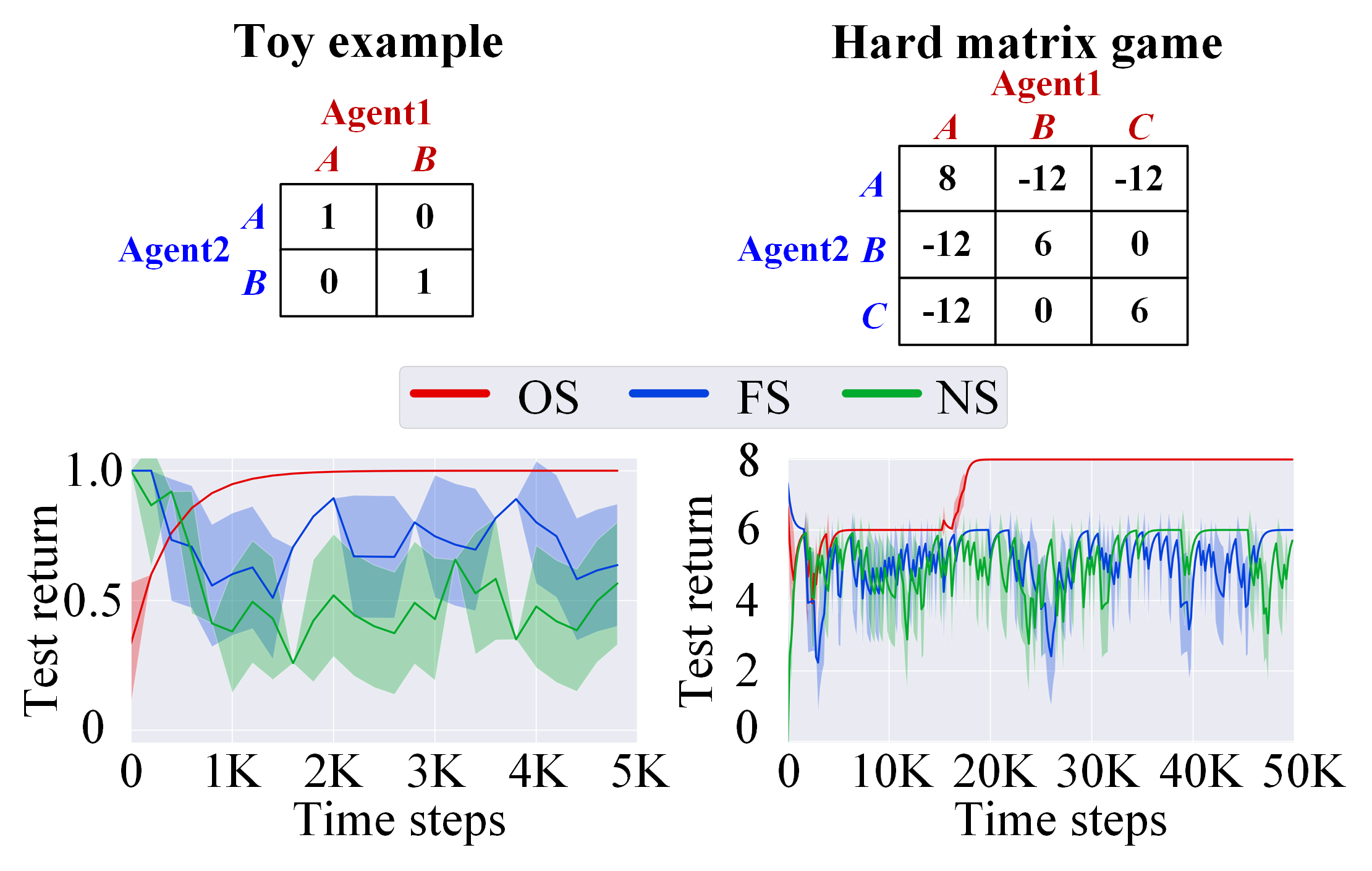} 
    \caption{Comparison of OS, FS, NS in matrix games.} 
    \label{fig:matrix} 
\end{figure}
Assuming that the intentions of Agent $1$ and Agent $2$ are $\widetilde{u}_1=\pi(o_1) $ and $\widetilde{u}_2=\pi(o_2)$. 
In FS, agents fully share intentions and change their decisions to $u_1=\pi(o_1,\widetilde{u}_2)$ and $u_2=\pi(o_2,\widetilde{u}_1)$ based on the received messages.
The message deceiving would happen if $\widetilde{u}_1\neq u_1$ or $\widetilde{u}_2\neq u_2$. 
In this case, the receiver is misled by the deceptive intention message, leading to potential miscoordination if $Q(u_1,u_2)<Q(\widetilde{u}_1,\widetilde{u}_2)$. 
By contrast, the leader-follower architecture does not suffer from message deceiving. 
The leader Agent $1$ selects its action independently $u_1(o_1)$ and propagates 
it to the follower Agent $2$, where Agent $2$ makes a decision $u_2=\pi(o_2,u_1)$ and does not affect the leader's choice.

We further investigate the performance of different intention sharing patterns in one-step cooperative matrix games involved two agents, where Agent $1$ and
Agent $2$ perform actions and receive a global reward according to the payoff matrix. The learning curves with the full exploration $\left(\varepsilon=1\right)$ is
shown in Figure \ref{fig:matrix}. We can see that OS can stably solve the tasks. NS and FS fluctuate significantly and fail to find the optimum solutions,
indicating that NS and FS suffer from training instability due to the non-stationarity and the message deceiving.

\section{Methodology}

\begin{figure}[t]
    \centering 
    \includegraphics[width=1.0\columnwidth]{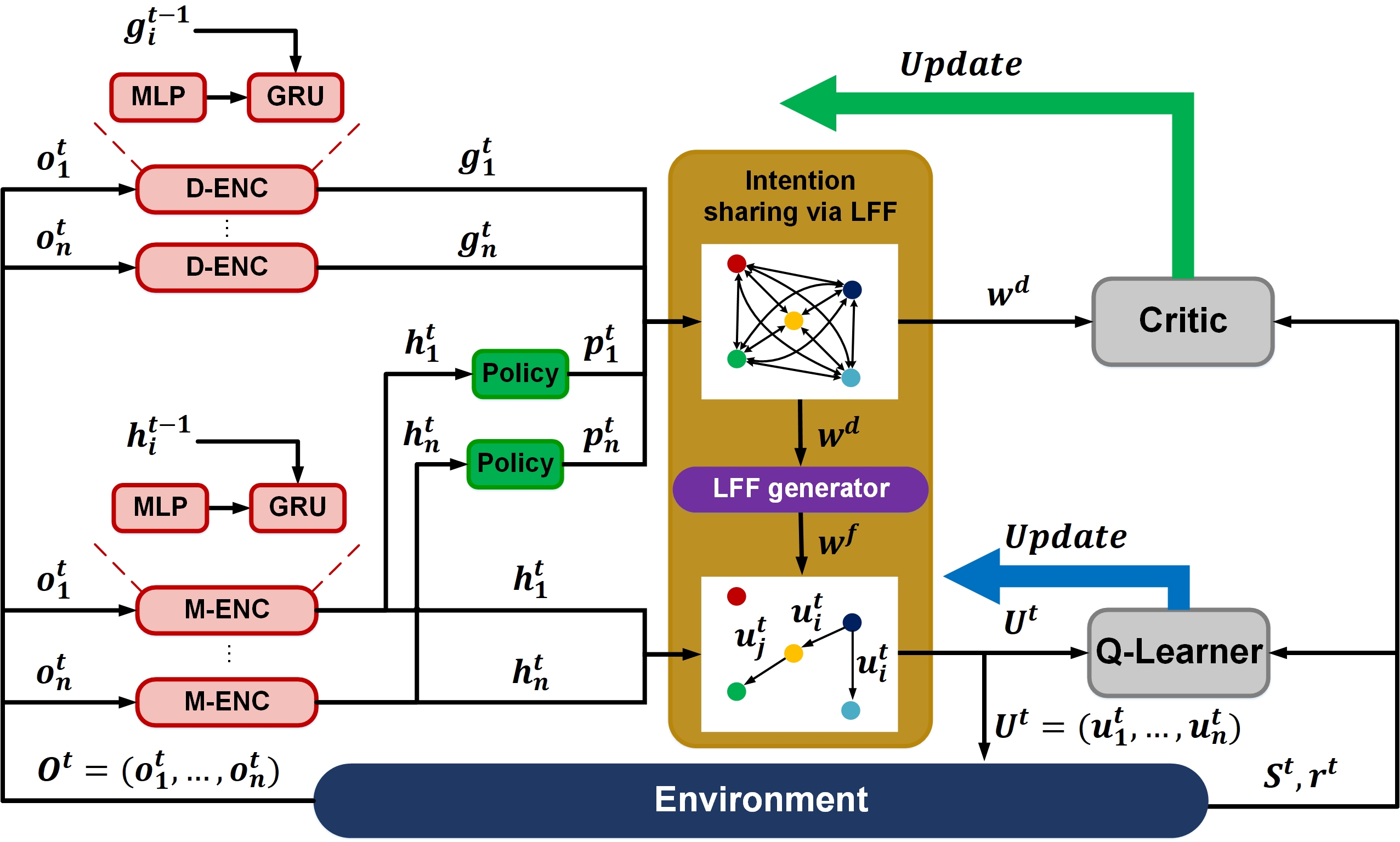} 
    \caption{Architecture of intention sharing via leader-follower forest.} 
    \label{fig:forward} 
\end{figure}
We consider a partially observable environment where $n$ agents cooperate to solve a task. 
As shown in Figure \ref{fig:forward}, each agent $i$ receives a local observation $o^t_i$ at time step $t$.
The local observation is encoded into a observation feature $h^t_i$ by a message encoder (M-ENC) and a dependency feature $g^t_i$
by a dependency encoder (D-ENC), respectively. 
The intention for each agent $i$ is $\widetilde{u}_i^t=f(h^t_i)$.
The dependency matrix $w^d$, which represents the interdependency between agents, is obtained according to $g^t_i$ and $\widetilde{u}_i^t$. 
Based on $w^d$, the hierarchical relationship $w^f$ is built through the leader-follower forest generator. 
During communication, agents share their intentions followed the order based on $w^f$, i.e., from the root node to the leaf node.
To be detailed, each agent $i$ makes the decision $u^t_i=\pi(h^t_i,c^t_i,x^t_i) $ based on its leaders' intentions, and then
send it to its followers,
where $c^t_i$ and $x^t_i$ denote the aggregated observation and intention messages from other agents, respectively.
Finally, agents perform joint actions $U=(u^t_1,...,u^t_n)$ and receive a global reward $r^t$ from the environment. 
The agent network is trained by linear value decomposition \cite{sunehag2018value} according to $r^t$. 
In addition, we model the task of finding out the optimal observation-dependent leader-follower forest as a reinforcement learning problem,
where the action space is defined as the dependency matrix $w^d$.
The forest is trained by DDPG \cite{lillicrap2016continuous} based on a communication reward $r^t_{c}$, which reflects the improvement of the decisions.
The pseudo-code is shown in Appendix.
In IS-LFF, the intentions flow from the root nodes to the leaf nodes, allowing agents to receive intentions from the parent node and their ancestors.
Consequently, the forest makes full use of information and guarantees one-sided intention sharing. 

\subsection{Intention sharing via leader-follower forest }

This subsection introduces intention sharing via leader-follower forest (IS-LFF) to dynamically construct the hierarchical relationship between agents and achieve one-sided intention 
sharing in multi-agent communication. The leader-follower forest (LFF) is defined as a set of directed rooted trees $F=\left(T_1,T_2,\ldots,T_n\right)$. Each tree is described as $T_i=\langle N,E \rangle$, where
$N$ and $E$ denote the set of nodes and directed edges, respectively. Each node represents an agent, and edges describe the leader-follower relationships between agent pairs. LFF constrains the intentions to
flow from the leaders to their followers through the directed edges, preventing message deceiving and alleviating non-stationarity during the communication.

We use a soft attention module to learn the interdependencies between agents. For agent $i$, the observation feature $h_i$ and the
intention $\widetilde{u}_i$ are encoded into a query vector $q_i\in\mathbb{R}^{d_k}$ and a key vector $k_i\in\mathbb{R}^{d_k}$, 
where $d_k$ denotes the dimension of the vector, and we omit time step $t$ from the notation for simplicity. Then the agent $i$'s 
dependencies on other agents are formulated as,
\begin{equation}
    \mu_i=softmax\left[\frac{q_i^{\top} k_1}{\sqrt{d_k}},\frac{q_i^{\top} k_2}{\sqrt{d_k}},\ldots,\frac{q_i^{\top} k_n}{\sqrt{d_k}}\right]
    \label{softmax} 
\end{equation}
Note that self-attention is not required because the agent would not change its decision without any intention messages from others,
i.e., $w_{ii}^d=0$. We use these dependency weights to form an $n\times n$ dependency matrix $w^d$ with $w^d_{ij}=\mu_{ij}$. 
In this way, a flow network is obtained where each edge represents the value of the dependency flow.
The value for each agent $i$ is defined as the difference between the ingoing and outgoing flows,
\begin{equation}
    \rho_i=-\sum_{\left(i,u\right)\in E} w_{iu}^d+\sum_{\left(v,i\right)\in E} w_{vi}^d
    \label{dv}
\end{equation} 
where $u$ and $v$ denote agents other than the given agent $i$. Since we apply soft attention to generate dependency weights 
and the self-attention is not used, we have $\sum_{\left(i,u\right)\in E} w_{iu}^d=1$.
As a result, the agent with a greater $\rho_i$ is more likely to locate at a higher level.

\begin{algorithm}[tb]
    \caption{Leader-follower forest generator}
    \label{alg:lff}
    \textbf{Input}: A dependency matrix $w^d$\\
    \textbf{Output}: A leader-follower forest $w^f$\\
    \begin{algorithmic}[1] 
    \STATE Initialise an agent set $N=(a_1,\ldots,a_n)$, a new agent set $N_{new}=()$.
    \STATE Calculate $\rho_i$ for each agent based Equation \ref{dv}.
    \IF {Use Top(k) selection}
    \STATE Include agents of the top $k$ value of $\rho_i$ to $N_{new}$.
    \ELSE
    \STATE Include the agent of the greatest value of $\rho_i$ to $N_{new}$.
    \ENDIF  
    \WHILE{$N_{new}\neq N$}
    \STATE Find the agent $i$ has the edge of maximum weight $w_{ij}$, where $i\in N_r=\left(N\backslash N_{new}\right)$ and $j\in N_{new}$.
    \IF {Use Top(k) selection}
    \STATE Include $i$ to $N_{new}$, restore $e_{ij}$.
    \ELSE 
    \STATE Find the maximum weight $w_{ir}$ of edges outgoing from the agent $i$.
    \IF {$w_{ij}=w_{ir}$}
    \STATE Include $i$ to $N_{new}$, restore $e_{ij}$.
    \ELSE
    \STATE Include the agent in $N_r$ of the greatest value of $\rho_i$ to $N_{new}$.
    \ENDIF
    \ENDIF
    \ENDWHILE
    \STATE Build $w^f$ based on recorded edges.
    \STATE \textbf{return} $w^f$.
    \end{algorithmic}
\end{algorithm}

Inspired by Minimum Spanning Tree \cite{prim1957shortest}, we propose the LFF generator to build the leader-follower relationship between agents.
First, we initialise an adjacency matrix $w^a$ and add the root node of the greatest $\rho_i$ to a new set $N_{new}$.
Next, we choose a candidate node $i$ has the edge of the maximum flow value $w_{ir}$, where $i\in \left(N\backslash N_{new}\right)$ and $r\in N_{new}$. 
The candidate node will be included to $N_{new}$ if $w_{ir}$ is greater than any flow value outgoing from it. Meanwhile, the directed edge $e_{ir}$ will be recorded.
Otherwise, it indicates that there are no suitable leaders for the candidate node in $N_{new}$, and then we reject this node and construct another tree based on $\rho_i$.
We repeat the processes until all agents are included in $N_{new}$ and finally obtain the leader-follower relationship $w^f$.
Alternatively, if the number of groups is given, we choose top $k$ nodes in terms of $\rho_i$ as root nodes in the first step to simplify the forest architecture, which is called Top(k)
selection. Note that the rejection mechanism is disabled in this case. The details of the LFF generator is shown in Algorithm \ref{alg:lff}.

During the communication, agent $i$ receives messages $m_i=\left[h_{-i},u_{-i}\right]$, where $h_{-i}$ and $u_{-i}$ denote the observation messages and intentions from agents other 
than agent $i$, respectively. A self-attention module is used to integrate the observation messages for each agent, and the leaders' intentions
is obtained based on $w^f$. Finally, agent $i$ makes the decision $u_i$ according to its observation feature $h_i$ and the aggregated message $aggr_i$,
\begin{equation}
    u_i=\pi(h_i, aggr_i)=\pi(h_i, c_i, x_i)
\end{equation}
\begin{equation}
    c_i=\sum_{j\in {N}} w^s_{ij} h_j, x_i=\bigcup _{j\in {N}} w^f_{ij} u_j = \bigcup _{j\in {L(i)}} u_j
\end{equation}
where $L(i) $ is the set of leaders of agent $i$, $w^s_{ij}$ and $w^f_{ij}$ denote the attention weight of messages and the relationship between agent $i$ and $j$, respectively.
Finally, agent $i$ sends its intention $u_i$ to followers and keeps it unchanged in this decision step.

\subsection{Learning algorithm based on LFF}

IS-LFF is divided into two parts: (i) the agent network parametrised by $\theta_{a}$, consisting of an encoder, a self-attention module, and a decision module; (ii) the
leader-follower forest determined by $w^d$. Inspired by SchedNet \cite{kim2018learning}, we propose a two-stage learning algorithm to train LFF-IS.

For the agent network, the linear value decomposition \cite{sunehag2018value} is applied to deduce the contribution of each agent in the presence of only a group reward. 
Since followers make decisions based on their leaders' intentions in IS-LFF, we decompose the joint action-value function into dependent individual value functions, i.e., $Q_{tot}=\sum_{i=1}^n Q_i(h_i,aggr_i)$.
For each agent, the individual value function is conditioned on its leaders' actions, alleviating the environment non-stationarity due to the agents’ changing policies.

In addition, the forest controls the message flows and thus plays a vital role in the decision-making process. 
Finding the dependency matrix $w^d$ is an ill-posed problem if the leader-follower forest $w^f$ is given. To make full use of the state
information and improve training stability, we treat the value function $V(s,w^f) $ instead of $V(s,w^d) $ as a bias, 
where $s$ denotes the state information. We omit $f$ in $w^f$ for simplicity.
The loss function for the agent network is,
\begin{equation}
\begin{aligned}
    L(\theta_{a},\theta_{v})=\sum_{b=1}^{B}\sum_{t=1}^{T} [(y^t_b-Q_{tot}(o^t_b,h^{t-1}_b;\theta_{a})\\
    -V(s^t_b,w^t_b;\theta_{v}))^2]
    \label{qlearning}
\end{aligned}
\end{equation}
where $B$ denotes the batch sampled from replay buffer $R$, and $T$ denotes the length of the episode. $ y^t_b=r^t_b + \gamma (Q_{tot}(o^{t+1}_b,h^{t}_b;\theta_{a}')+V(s^{t+1}_b,w^{t+1}_b;\theta_{v}')) $
is the target value. $\theta_{a}'$ and $\theta_{v}'$ are the parameters of target networks, which are updated by periodically copying the parameters from the source network.

On the other hand, we design a communication reward $r_{c}$ and apply DDPG \cite{lillicrap2016continuous} to learn the leader-follower forest. 
First, inspired by VBC \cite{zhang2019efficient}, we use the variance between the largest and the second-largest action values to model the confidence 
of agent $i$'s decision. Therefore, an intrinsic reward for the improvement of decision confidence is formulated as,
\begin{equation}
    r_{int}^t=\sum_{i\in N}{[v_i(h_i^t,aggr_{i}^t)-v_i(h_i^t,0)]}
\end{equation}
where $v_i(h_i^t,c_{-i}^t) $ and $v_i(h_i^t,0)$ denote the confidence of agent $i$'s decisions before and after communication, respectively. 
We use $V(s^t_b,w^t_b)$ from Equation \ref{qlearning} to evaluate the performance of the forest. Therefore, the communication reward for the leader-follower forest is,
\begin{equation}
    r_{c}^t=V(s_b^t,w_b^t)+\alpha r_{int}^t
    \label{rins}
\end{equation}
where $\alpha$ denotes the coefficient of the intrinsic reward. We treat the dependency matrix $w^d$ as an action $a=\varphi(o)$, 
where $\varphi(o)$ is a mapping from the observations to the agents' dependencies on others. The dependency matrix is updated based on DDPG,

\begin{equation}
    L(\theta_{c})=\mathbb{E}_{o,s,a,o',s'\sim R}[(Q(s,a;\theta_{c})-y)^{2}] 
    \label{mse}
\end{equation}
\begin{equation}
    \nabla  _{\theta_d}J(\varphi) =\mathbb{E}_{o,s,a\sim R} [\nabla_{\theta_d}\varphi(a\vert o) \nabla_{a} Q(s,a;\theta_{c})\vert _{a=\varphi(o)}]
    \label{ddpg}
\end{equation}
where $y = r_{c} + \gamma Q(s',a';\theta_{c}^{'})\vert _{a'=\varphi^{'}(o')} $ is the target value, and
$\theta_{c}^{'}$ denotes the parameters of target critic network.

\subsection{Depth-bounded reconstruction}
In practice, the communication time may be strictly limited, leading to the depth boundary of the forest. In this situation, we reconstruct the forest according to the given depth boundary $d$, i.e.,
we prune out nodes located in layers deeper than $d$, and then reassign new parent nodes from the top $d-1$ layer for these isolated nodes layers based on the dependency matrix $w^d$. 
In this way, we convert the leader-follower forest to the depth-bounded forest to meet the communication time requirements.

\section{Related work}
Learning communication protocols to solve cooperative tasks is one of the desired emergent behaviours of agent interactions and has become an active area in multi-agent reinforcement
learning \cite{lazaridou2017multi, hernandez2019survey}. Several algorithms have been put forward to help agents share information through differential protocols in recent years, 
and we classify them into three categories. The first category focuses on effectively sharing observation information to solve partially observable problems. RIAL and DIAL \cite{foerster2016learning} use discrete 
communication channels to demonstrate end-to-end learning of protocols. By contrast, CommNet \cite{sukhbaatar2016learning} uses continuous channels, averaging
message vectors and sending them to each agent. IC3Net \cite{singh2018learning} controls when to communicate by a gate and can be applied to both cooperative and 
competitive scenarios. ATOC \cite{jiang2018learning} proposes a communication model based on the hard attention mechanism to learn when to communicate and integrate information from others. 
TarMAC \cite{das2019tarmac} applies a soft attention matrix to learn the relationship and aggregate messages based on the attention weights. VBC \cite{zhang2019efficient} enables the agents to send
communication requests and reply to others adaptively based on their confidence about local decisions, reducing unnecessary information exchange. SchedNet \cite{kim2018learning} schedules the agents by a 
weight generator, and only a few agents are allowed to broadcast messages. NDQ \cite{wang2019learning} uses message entropy and mutual information to shorten the message and reduce the uncertainty of
the receiver's state-action value. 

The second category models the relationships between agents by graphs and propagates messages through the edges. Graph Neural Network \cite{battaglia2018relational} learns message passing
over the edges and the joint state-action value, which requires additional losses and might not be feasible in practice. Coordination Graph \cite{guestrin2002coordinated}
is proposed to decompose the individual state-action value into a utility function and a payoff function based on the messages of connected agents. Deep
Coordination Graph \cite{bohmer2020deep} applies deep neural networks and parameter sharing to deal with large state and action spaces. However, the graph
is usually pre-defined in these methods. DICG \cite{li2021deep} uses the attention mechanism to learn the appropriate message-dependent coordination graph
structure with soft edge weights. GA-Comm argues \cite{liu2020multi} that soft attention makes the agent still rely on irrelevant agents' messages and proposes
a game abstraction mechanism to extract relationships.

The third category includes intentions into messages and allow agents to negotiate during communication, reducing the non-stationarity caused by the uncertainty of agents' policies. 
Some non-communication methods solve this problem by modelling others, i.e., predicting others' behaviours or goals \cite{raileanu2018modeling,rabinowitz2018machine,li2021dealing}.
By contrast, communication allows agents to share their intentions directly with others instead of inference. NeurComm \cite{chu2019multi} takes policy fingerprints into communication 
to reduce information loss and non-stationarity. Intention propagation \cite{qu2020intention} allows agents to spread out initial decisions to their neighbours, 
converging to the mean-field approximation of the joint policy. In IS \cite{kim2020communication}, the agents model the environment dynamics to predict their imaginary paths and
share intentions with others by an attention module. Although intentions are helpful in coordination, agents may change their decisions after communication simultaneously in these
methods. In this way, the shared intentions cannot reflect agents' final actions, leading to potential message deceiving and miscoordination.

The most related works to this paper are IS and GA-Comm. We consider that message deceiving may occur in full or attentional intention sharing. On the other hand,
the existing graph methods cannot generate leader-follower relationships to ensure one-sided intention sharing. In comparison, our approach uses trainable leader-follower 
forests to constrain intentions to flow through directed edges, avoiding message deceiving and improving the performance of joint policy. In addition, our method decomposes 
the joint state-action value function into dependent value functions, alleviating the environment non-stationarity.
\begin{figure}[t]
    \centering
    \includegraphics[width=1.0\columnwidth]{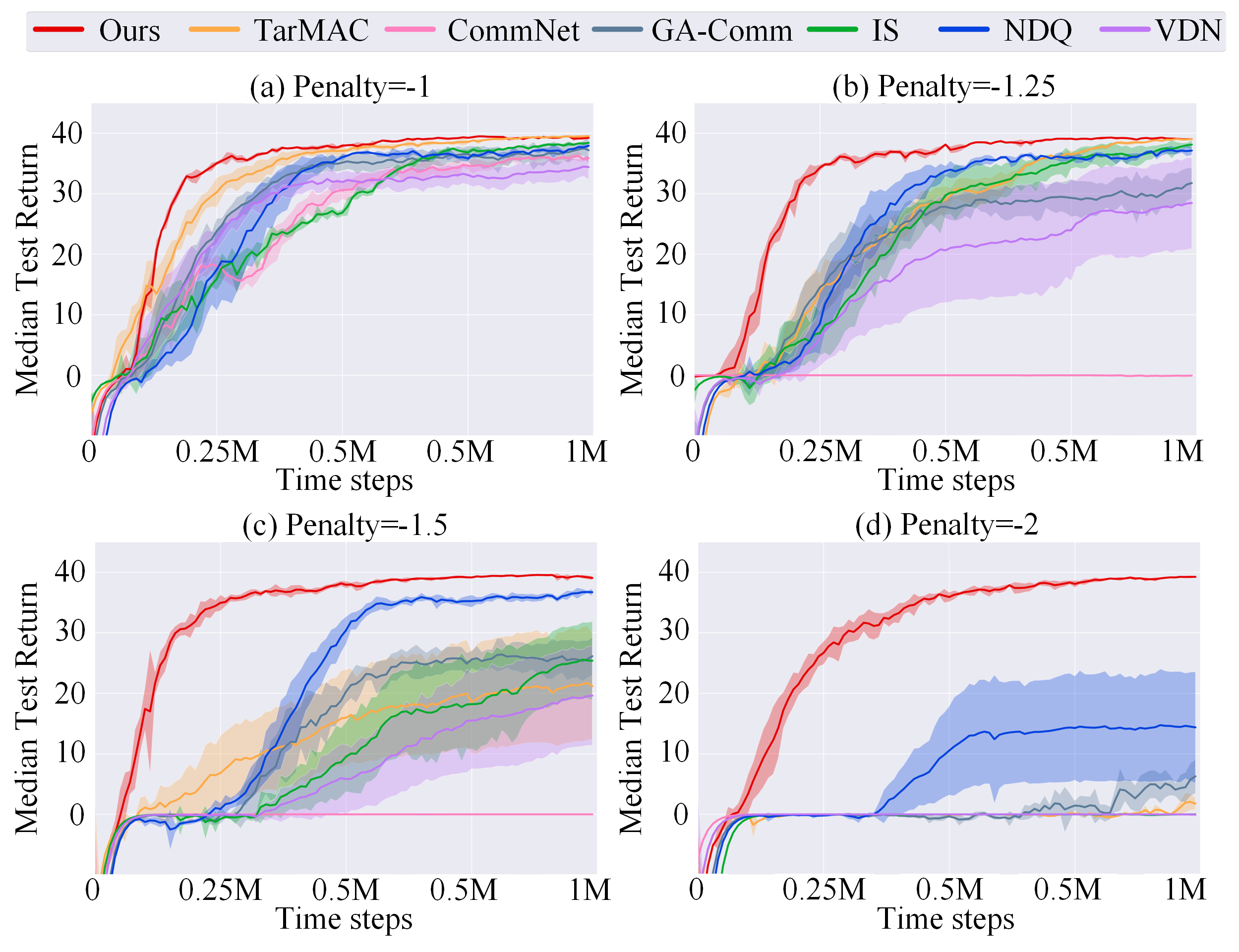} 
    \caption{Reward of IS-LFF against baselines during training on predator-prey with different penalties.}
    \label{pp}
\end{figure}

\section{Results}
We evaluate our proposed method on predator-prey and StarCraft II Multi-agent Challenge, which have recently become popular benchmarks in multi-agent reinforcement learning
under partially observable settings \cite{rashid2018qmix,samvelyan2019starcraft,son2019qtran,bohmer2020deep,zhang2019efficient,rashid2020weighted}. We compare the performance of LFF with a variety of state-of-the-art communication algorithms, 
including CommNet \cite{sukhbaatar2016learning}, TarMAC \cite{das2019tarmac}, NDQ \cite{wang2019learning}, GA-Comm \cite{liu2020multi}, and IS \cite{kim2020communication}. 
For fairness, we combine our baselines with VDNS since most of them do not deal with credit assignment problems (except for NDQ).
We also conduct ablation studies to answer the following questions: (i) How does our method differ from intention sharing based on pre-defined 
topologies or existing graph methods? (ii) How do the number of groups $k$ and the depth boundary $d$ influence the communication performance and cost time?
All results are averaged over five seeds, and the standard error of means are shaded. More details about the implementation of baselines and our method are included in Appendix. 

\subsection{Predator-Prey}
The partially observable predator-prey environment consists of $8$ predators and $8$ prey in a $10\times10$ grid world. The predator can only observe a $5\times5$ sub-grid around it. A 
successful catch is defined as that two adjacent agents perform catch operation simultaneously, which is rewarded with a global reward $r=10$. The miscoordination, i.e., only one 
agent attempts to catch, is punished by a penalty $p\le0$. It is proved that the stronger penalty causes failures in many non-communication algorithms \cite{bohmer2020deep, rashid2020weighted}.

Figure \ref{pp} illustrates the median test return for different miscoordination penalties over five runs. We can see that CommNet, TarMAC, GA-Comm become unstable
and even fail with the increase of penalty. In particular, CommNet fails to solve the task when $p\le-1.25$, suggesting that the noisy and redundant communication would impair coordination.
NDQ uses mutual information to reduce the non-stationarity and thus can learn the task. However, we can see that significant standard errors of NDQ when $p=-2$, suggesting
that the additional entropy loss is not feasible and reliable in this task. 
IS also fails to solve the task even though it propagates the intention through a soft-attention module.
It is difficult for soft-attention modules to learn a forest and achieve one-sided intention sharing because
they use the softmax activation function to obtain the relative weights of the relationship between agents.
In IS-LFF, agents make decisions based on their leaders' intentions, i.e., the follower agent would execute catch only if it receives the message that its
leader decides to do so. Therefore, our algorithm learns quickly and reliably compared to the baseline algorithms in all penalty settings.

\subsection{StarCraft Multi-Agent Challenge (SMAC)}
This paper focuses on the decentralised micromanagement problem in SMAC, which is based on the popular real-time strategy game StarCraft II.
More details can be found in Appendix.
We compare the performance of our method and baselines on
two hard maps (\texttt{2c\_vs\_64zg} and \texttt{3s\_vs\_5z}) and two super hard maps (\texttt{3s5z\_vs\_3s6z} and \texttt{corridor}) because the existing non-communication methods 
do not perform very well in these scenarios \cite{samvelyan2019starcraft}\footnote{We use \texttt{SC2.4.6.2.69232} (the same version as VDN and QMIX) instead of 
\texttt{SC2.4.10}. Performance is \textbf{not} comparable across versions.}. 

\begin{figure}[t]
    \centering
    \includegraphics[width=1.0\columnwidth]{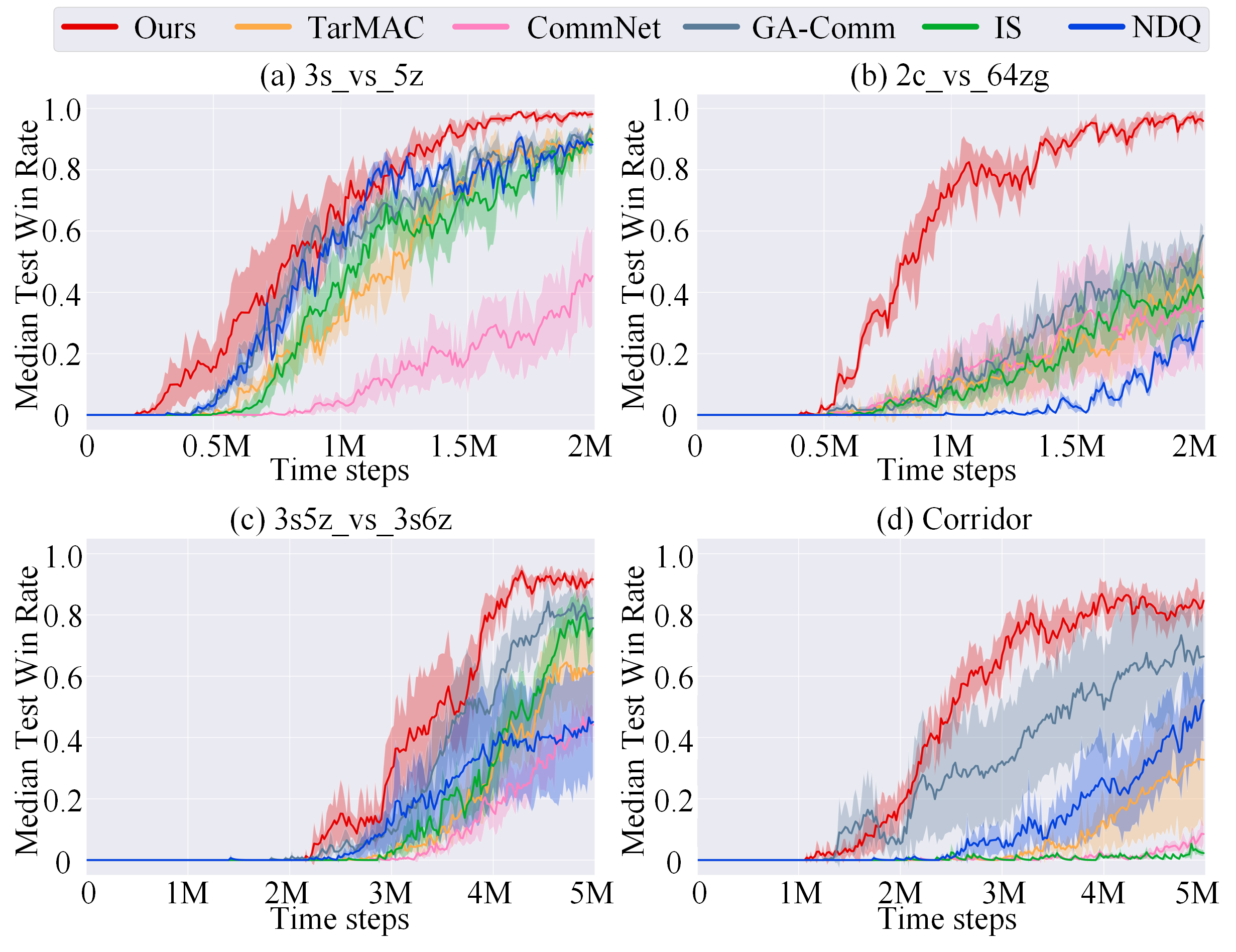} 
    \caption{Median test win rate of IS-LFF against baselines on SMAC tasks.}
    \label{smac_p}
\end{figure}

Figure \ref{smac_p} shows the median win rate for our proposed algorithm and baselines. We use an $\epsilon$-greedy policy in which $\epsilon$ decreases from 1 to 0.05
over 0.5 million timesteps in \texttt{3s\_vs\_5z} and \texttt{2c\_vs\_64zg}, and over 1 million timesteps in \texttt{3s5z\_vs\_3s6z} and \texttt{corridor} for all algorithms.
We can see that LFF significantly outperforms baselines, indicating that sharing intention via the leader-follower forest can improve learning speed and achieve better coordination 
in practice.
To further demonstrate the effectiveness of our method, 
we also combine communication protocols with QMIX \cite{rashid2018qmix}, which is a popular credit assignment algorithm. In this setting,
LFF-IS still outperforms our baselines. The results can be found in Appendix.

\subsection{Ablation}
Finally, we present the results of ablation studies to demonstrate the effectiveness of our algorithm. First, we investigate the necessity of learning observation-dependent LFF
compared to the pre-defined topologies and existing graph methods. We choose two pre-defined topologies, i.e.,
line and random patterns, where agents share their intentions one by one or followed the order given by a random forest, respectively.
We also design two intention-sharing schemes based on existing graph methods, i.e., attentional intention sharing (AIS) and game abstraction 
intention sharing (GA-IS). AIS allows each agent to propagate intentions via a soft attention module and change their strategies based on received messages.
In GA-IS, we integrate intention information into messages based on GA-Comm. To ensure fairness, we set the depth boundary $d$ equals the number of communication rounds
in AIS and GA-IS, which is 3 in all scenarios.

\begin{figure}[t]
    \centering
    \includegraphics[width=1.0\columnwidth]{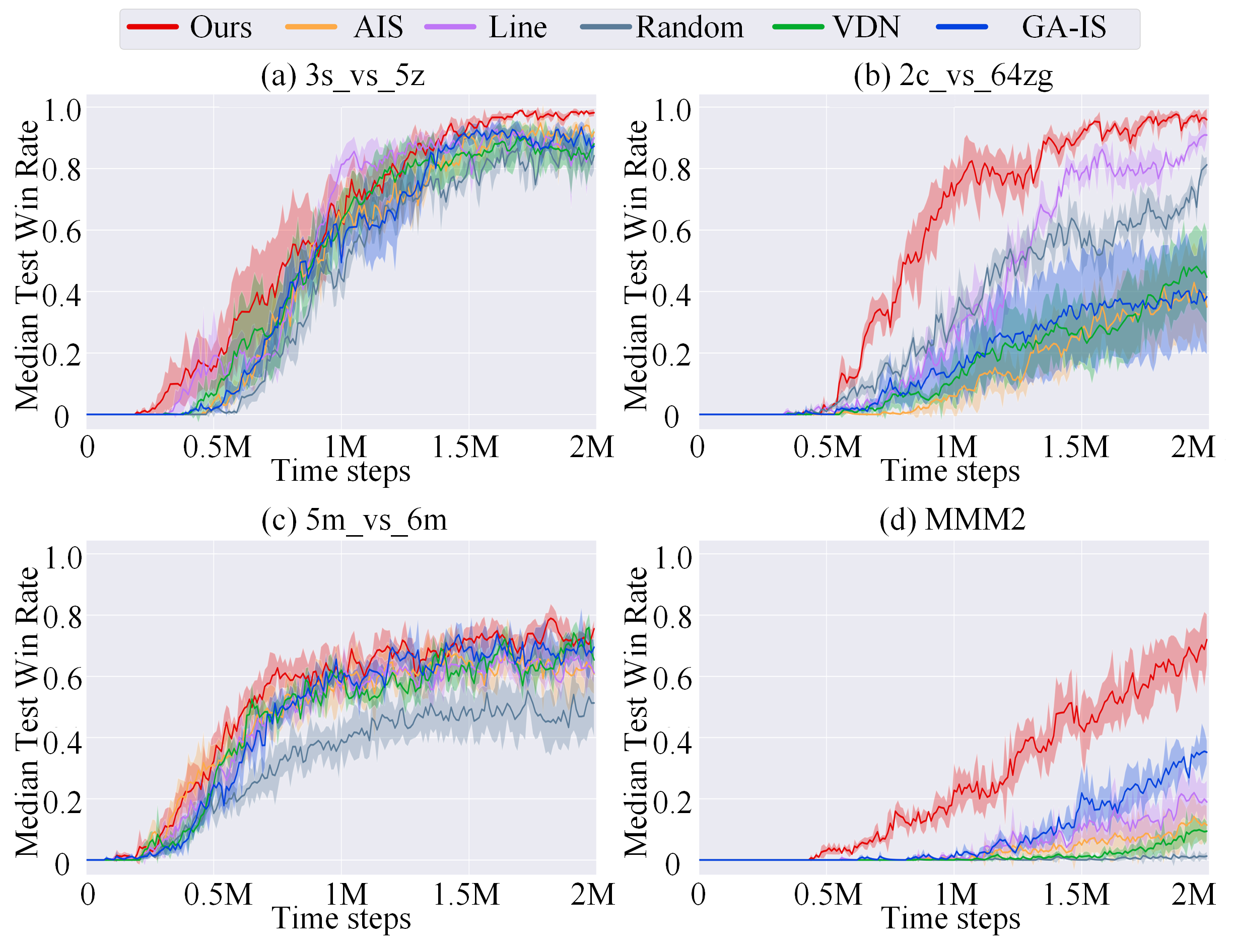} 
    \caption{Comparison of winning rates between our method and other topologies.}
    \label{fig:nofs}
\end{figure}

Figure \ref{fig:nofs} demonstrates that our method outperforms the intention sharing through pre-defined topologies or graphs generated by existing methods. 
The poor performance of AIS and GA-IS indicates that the message deceiving is widespread, and
it is unlikely to achieve one-sided intention sharing based on attention modules.
Line and random patterns benefit from one-sided information sharing and improve the performance compared to VDN.
We can see that the line pattern outperforms the random because it keeps leaders' intentions and provides sufficient information for followers.
However, the line pattern is time-consuming and cannot represent the proper relationship between agents.
By contrast, our method learns faster and achieves the best performance, verifying the effectiveness of intention sharing 
via leader-follower forest.

\begin{figure}[t]
    \centering
    \includegraphics[width=1.0\columnwidth]{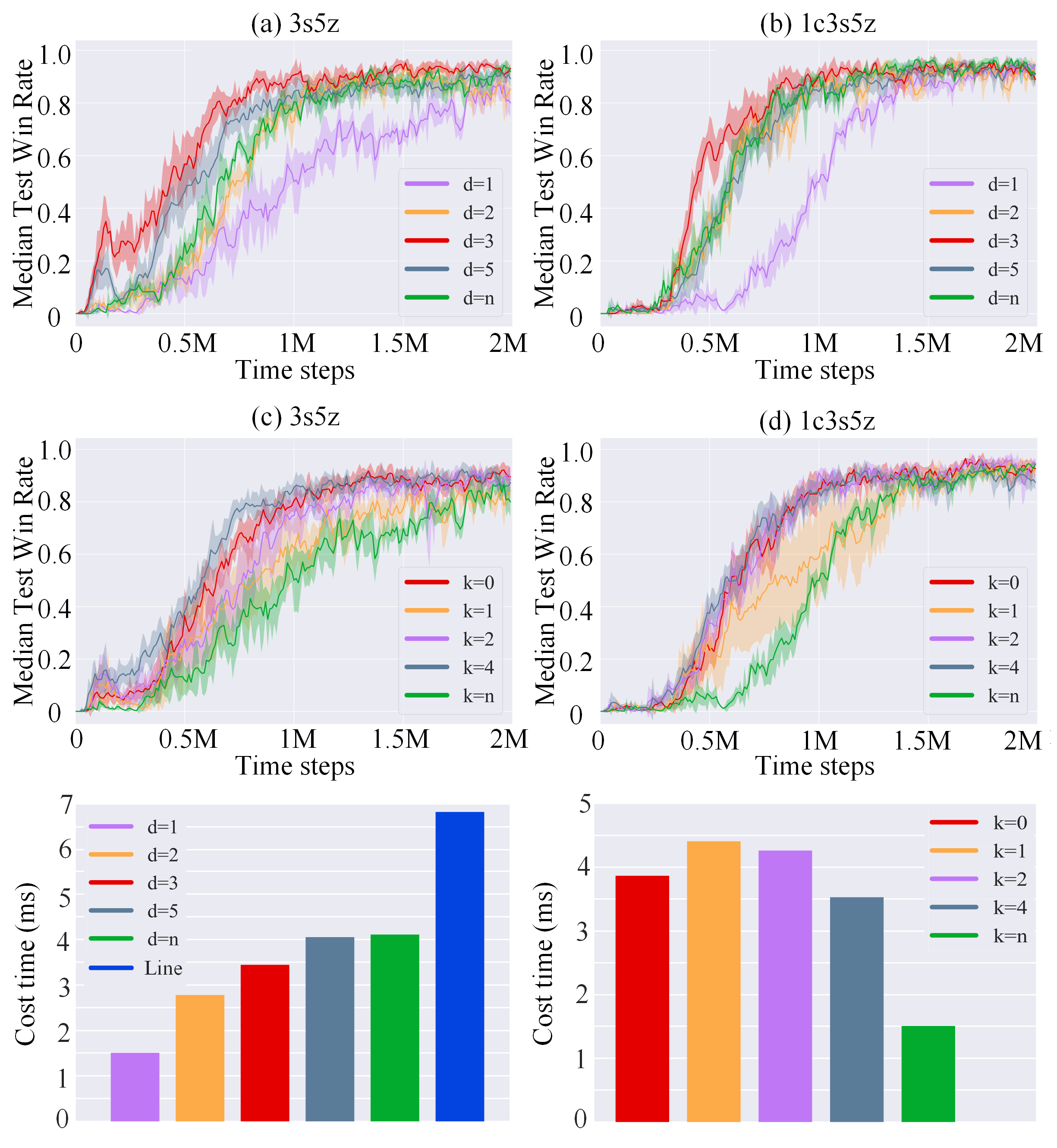} 
    \caption{The influence of the number of trees $k$ and the depth boundary $d$. $k=0$ represents the leader-follower forest without Top(k) selection. $n$ denotes the number of agents in the scenario.}
    \label{fig:hyper}
\end{figure}

Second, we compare the performance and cost time of LFF with different trees $k$ and depth boundaries $d$ in Figure \ref{fig:hyper}.
We can see the communication time reduces considerably with decreasing $d$.
However, if $d$ is too low, the leader-forest tree degenerates to the no intention sharing structure, leading to message
deceiving and potential miscoordination. 
The reasonable $k$ and $d$ balance the performance and communication time 
because they reduce the variety of leader-forest trees and thus cut exploration time, 
leading to faster converging and better performance.
In practice, we can determine the optimal values for $k$ and $d$ by the trial-and-error method based on the given task-specific constraints.

\section{Conclusion and future work}

This paper proposes intention sharing via leader-follower forest, allowing agents to share their intentions through observation-dependent
leader-follower forests to solve cooperative tasks in multi-agent systems. The leader-follower forest is built based on the interdependency
between agents, ensuring one-sided intention sharing and eliminating message deceiving. To train the proposed communication scheme, we design a two-stage learning algorithm:
DDPG with a communication reward for the forest and VDN for the agent network. In addition, we propose the depth-bounded LFF to reduce communication
time. The results on predator-prey and SMAC show that IS-LFF benefits from one-sided intention sharing and
outperforms state-of-the-art communication algorithms. The ablation studies demonstrate that IS-LFF outperforms intention sharing based on 
pre-defined topological structures and other graph methods. The number of trees and the depth boundary affect the performance and the
communication time of LFF, where the optimal values are task-specific. 

We apply DDPG to learn the leader-follower forest, which is effective but cannot guarantee to find
the optimal structure of the forest. 
The agents could learn directed acyclic graphs more flexibly and efficiently by introducing reinforcement
learning algorithms in combinatorial optimisation. We would investigate it in future work.

\appendix
\bibliography{aaai22}
\section{Appendix}
\subsection{Pseudo Code}
\label{Intention sharing via leader-follower forest}
\begin{algorithm}[h]
    \caption{IS-LFF}
    \label{alg:IS-LFF}
    \begin{algorithmic}[1] 
    \STATE Initialise actor parameters $\theta_{a}$, LFF parameters $ \theta_{d}$, 
    critic parameterss $ \theta_{c}$, target networks parameters $ \theta_{a}^{'}$, $ \theta_{d}^{'}$, $ \theta_{c}^{'}$,
    and a replay buffer $R$.
    \FOR {episode=$1,2,...,M$}
    \FOR {$t=1,2,...T$}
    \STATE Obtain the observation encoding $h_{i}^{t}$ and initial decision $ \widetilde{u}_i^t $ for each agent.
    \STATE Calculate dependency matrix $w^d$ and build LFF $w^f$ based on LFF generator.
    \STATE Each agent $i$ selects an action $u_i^{t}$ based on the aggregated message $aggr_{i}^t $, and propagates
    it to followers through directed edges in LFF.
    \STATE Execute $U^t=(u_1^t,...,u^t_n)$, obtain next observation $o_{i}^{t+1}$ for each agent and a global reward $r^{t} $.
    \STATE Store transition $(o^{t},s^{t},a^{t},r^{t},o^{t+1},s^{t+1})$ in $R$.
    \ENDFOR
    \STATE Update $\theta_{a}$ by minimise the loss presented in Equation $5$.
    \STATE Update $\theta_{c}$ by minimise the loss presented in Equation $8$.
    \STATE Update $\theta_{d}$ using sampled policy gradients based on Equation $9$.
    \STATE Update $ \theta_{a}^{'}$, $ \theta_{c}^{'}$, and $ \theta_{d}^{'}$ periodically copied 
    from $ \theta_{a}$, $ \theta_{c}$, and $ \theta_{d}$.
    \ENDFOR
    \end{algorithmic}
\end{algorithm}

\subsection{Additional results}
\label{additional}
\begin{figure}[t]
    \centering
    \includegraphics[width=1.0\columnwidth]{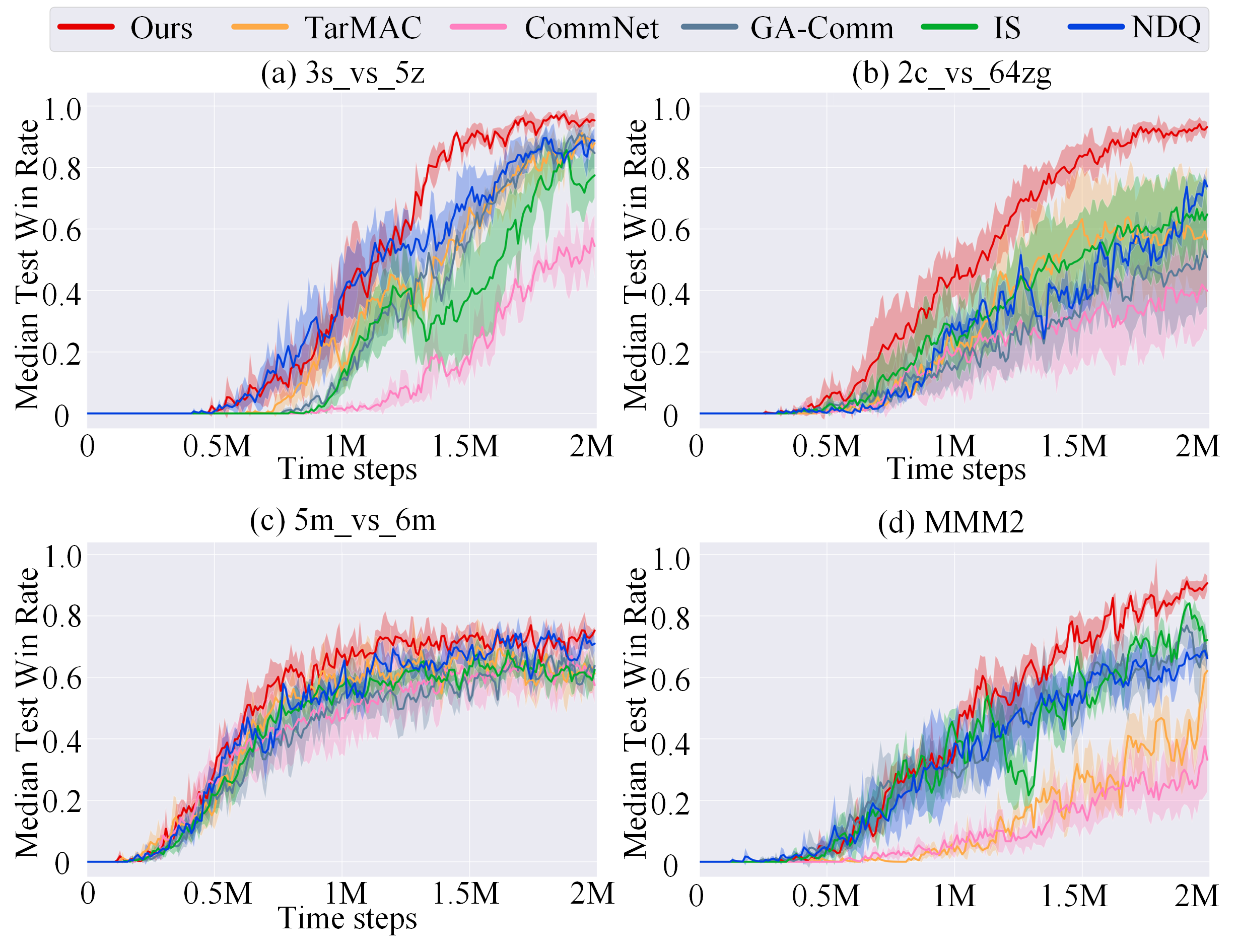} 
    \caption{Comparison of winning rates between baselines and our method in QMIX version.}
    \label{smac_qmix}
\end{figure}

\begin{figure}[t]
    \centering
    \includegraphics[width=1.0\columnwidth]{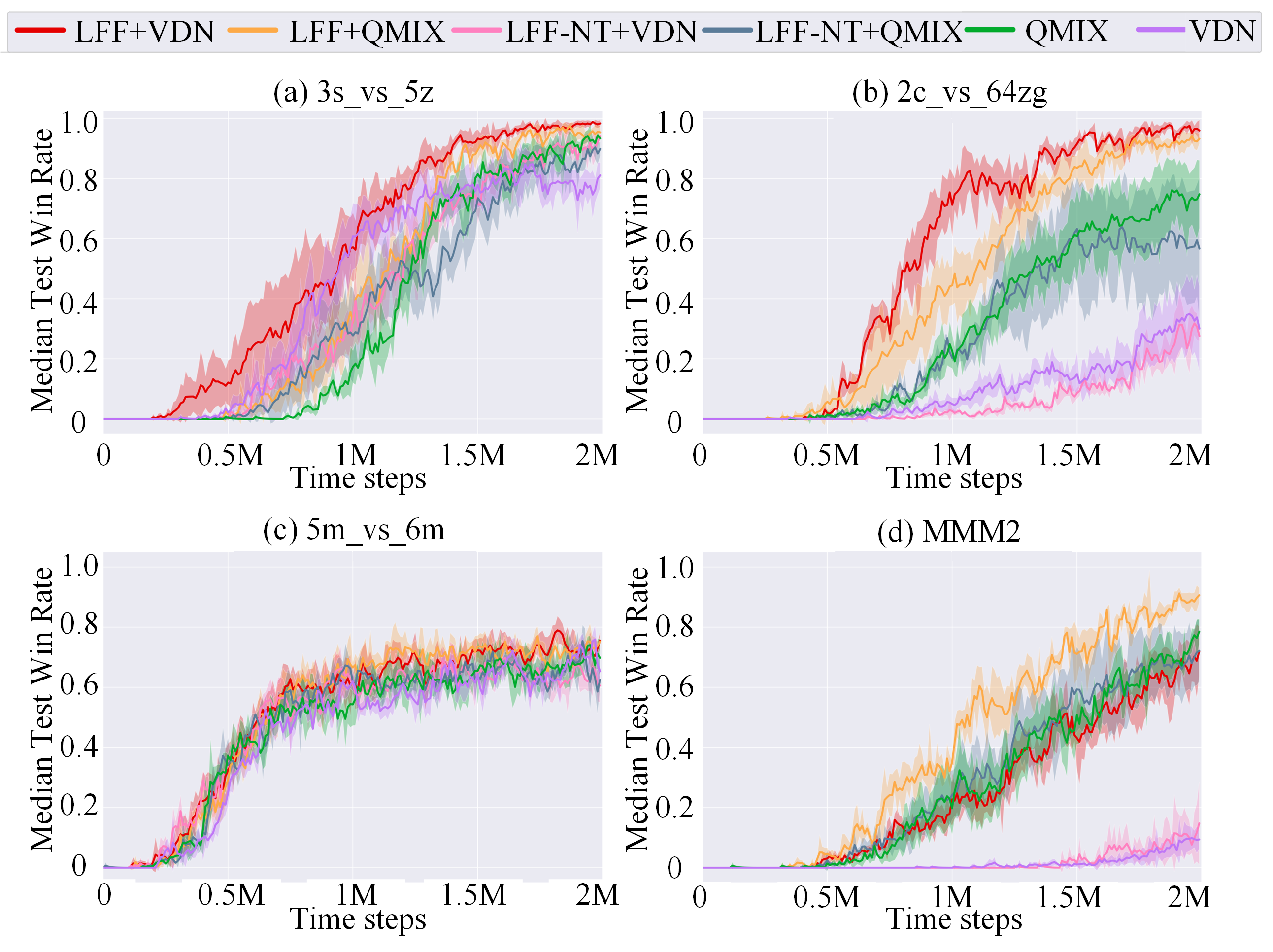} 
    \caption{Comparison of winning rates of method combined with VDNS and QMIX. LFF-NT+QMIX represents the IS-LFF without the leader-follower forest.}
    \label{vdn_qmix}
\end{figure}
We compare the performance of communication methods combined with another credit assignment algorithm, i.e., QMIX. In Figure \ref{smac_qmix}, we can see that QMIX
helps our baselines achieve better winning rates compared to the VDNS. However, LFF still outperforms them, suggesting the effectiveness of intention sharing via leader-follower forest.
Figure \ref{vdn_qmix} shows that LFF+VDN improves the performance of VDNS and LFF-NT+QMIX significantly, which even learns faster than LFF+QMIX in \texttt{3s\_vs\_5z} and \texttt{2c\_vs\_64zg}.
However, LFF+QMIX achieves the best performance in \texttt{MMM2}, which indicates that more hyperparameters are required in complex scenarios involved a great number of agents.

\begin{figure}[t]
    \centering
    \includegraphics[width=1.0\columnwidth]{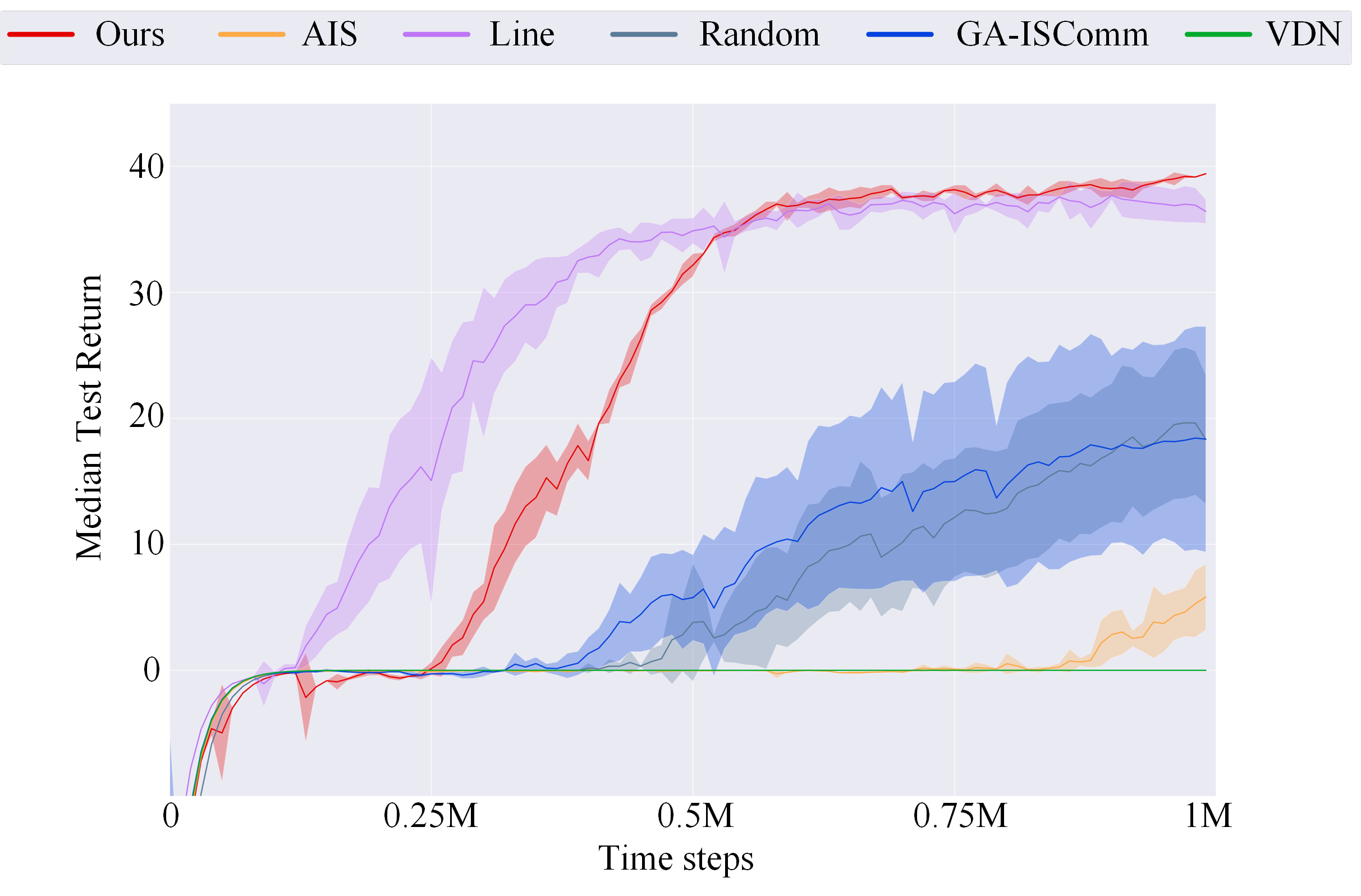} 
    \caption{Reward of IS-LFF against other topologies on predator-prey with penalty=$-2.25$.}
    \label{pp^topo}
\end{figure}

On the other hand, we also investigate the influence of different topological architectures for one-sided intention sharing on predator-prey. Figure \ref{pp^topo} shows the results of the
comparison between IS-LFF and other topologies when the penalty for miscoordination is $-2.25$. Random pattern and GA-ISComm perform poorly (we can see a considerable fluctuation), and AIS and
VDN fail to solve the task. By contrast, the line pattern and IS-LFF learn steadily and can solve the task. We also can see that the line learns faster than IS-LFF in the early stage, suggesting the 
difficulties in learning a certain observation-dependent relationship between agents. However, the final performance of our method is greater than the line, indicating that learning dynamically
generated leader-follow forests is necessary to achieve better coordination.

\subsection{StarCraft II environment}
\label{smac}

StarCraft Multi-Agent Challenge is a partially observable reinforcement learning benchmark. An individual agent with parameter sharing 
controls each allied unit, and a hand-coded built-in StarCraft II AI controls enemy units. The difficulty of the game AI is set to the ``very difficult'' level. 
At each time step, agents have access to their local observations within the field of view. The feature vector 
contains attributes of both allied and enemy units: \texttt{distance}, \texttt{relative\ x}, \texttt{relative\ y},
\texttt{health}, \texttt{shield}, and \texttt{unit\_type}. In addition, agents can observe the last actions of allied units and the terrain
features surrounding them.

The global state vector includes the coordinates of all agents relative to the centre of the map and other features 
present in the local observation of agents. In addition, the state stores the energy of Medivacs, the cooldown of the rest 
of the allied units and the last actions of all agents. Note that the global state information is only available
to agents during centralised training. All features both in state and local observations are normalised 
by their maximum values.

After receiving the observations, each agent is allowed to take action from a discrete set which consists of 
\texttt{move[direction]}, \texttt{attack[enemy\_id]}, \texttt{stop} and \texttt{no-op}. Move direction includes
north, south, east, and west. Note that the dead agents can only take \texttt{no-op} action while live agents cannot. 
For health units, Medivacs use \texttt{heal[agent\_id]} actions instead of \texttt{attack[enemy\_id]}. The maximum 
number of actions varies between $7$ and $70$ depending on different scenarios. Note that agents can only perform the
\texttt{attack[enemy\_id]} action when the enemy is within its shooting range. 

At each time step, agents take actions and receive a positive global reward based on the total damage dealt on
the enemy units. In addition, they can receive an extra reward of $10$ points after killing each enemy unit and 
$200$ points after killing all enemy units. The rewards are scaled to around $20$ so that the maximum cumulative reward achievable
in each scenario.

\subsection{Experimental setup}
\label{baseline}
The detailed implementations of the communication baselines and our method are as follows. We use RMSProp with a learning rate of $5\times 10^{-4} $ and $\alpha=0.99 $, buffer size $5000$,
mini-batch size $32$, discount factor $\gamma=0.99 $ for all algorithms. We combine all baselines with VDNS (unless 
specified otherwise), which uses a $64$-dim hyper network to generate a state-dependent bias to deal with the credit 
assignment problem. The dimension of each agent’s encoder hidden state is set to $64$.

TarMAC, GA-Comm, IS, and our method share the same soft attention network to aggregate messages. The message signature/query dim is
$32$, and the message dim is $64$. We also use another attention network, which has the same number of parameters as the soft one, to build 
the dependency matrix as the input for the LFF module. The architecture of the DDPG critic is a feed-forward network with $3$ hidden 
layers of dim $\{ 128,64\}$ and ReLU non-linearities. We do not use Top(k) selection, the depth boundary is set to $3$, and the relative weight of 
intrinsic reward is set to $0.1$ in all scenarios.

Note that we use a Bi-GRU of dim $64$ in GA-Comm to achieve hard attention.

In NDQ, we use two $20$-dim hyper networks with ReLU activation for the posterior estimator and do not cut messages in the training process.

For multi-round communication, CommNet and TarMAC use the same communication round as $3$.

In IS, the length of the imagined trajectory is set to $3$.

In AIS, we apply a soft attention matrix similar to TarMAC to help agents learn the targets they need to receive or send intention information. 
The number of communication rounds for AIS is set to $3$. In GA-IS, we use the same architecture as GA-Comm and only add intentions into messages.  

\end{document}